\documentstyle[12pt,epsf]{article}
\newcommand{\sptwo}{1.4}

\newcommand{\doublespace}{\edef\baselinestretch{\sptwo}\Large\normalsize}

\begin{document}
\doublespace

$~$
\vspace{55pt}

\begin{center}
{\bf TEN YEARS OF THE FIFTH FORCE$^{*\dagger}$}\\
Ephraim Fischbach and Carrick Talmadge\\
Physics Department, Purdue University, West Lafayette, IN  47907-1396 USA\\
\end{center}
\vspace{3cm}

\begin{center}
{\bf ABSTRACT}
\end{center}

\begin{quote}
The suggestion in 1986 of a possible gravity-like ``fifth" fundamental
force renewed interest in the question of whether new macroscopic
forces are present in nature.  Such forces are predicted in many
theories which unify gravity with the other known forces, and
their presence can be detected by searching for apparent
deviations from the predictions of Newtonian gravity.  We review the
phenomenology behind searches for a ``fifth force", and present a
summary of the existing experimental constraints.
\end{quote}

\vspace{225pt}

\noindent
\underline{$~~~~~~~~~~~~~~~~~~~~~~~~~~~~~~~~~~~$}

\noindent
$^*$Work supported in part by the U.S. Department of Energy.

\noindent
$^\dagger$Invited talk presented at the XXXI Recontres de Moriond, 20-27 January, 1996.

\noindent
To be published in the conference proceedings.

\pagebreak

This year marks the 10th anniversary of the ``fifth force"
hypothesis --- the suggestion that there exists in nature a
new intermediate range force similar to gravity, and co-existing
with it [1-9].  Much of the work carried out during this period
has been reported at Moriond, and so it is appropriate to use
the occasion of this year's Moriond Workshop to review
what we have learned during the past decade.

In the simplest models, a ``fifth force" would arise from
the exchange of a new ultra-light boson which couples to
ordinary matter with a strength comparable to gravity.
There are numerous theories of physics at the Planck scale
which predict the existence of such ultra-light bosonic
fields [3-6], whose effect is to modify the expression for
the interaction energy $V(r)$ for two point masses $m_1$
and $m_2$:
$$  V(r) = \frac{-G_\infty m_1m_2}{r} (1 + \alpha e ^{-r/\lambda}).
   \eqno{(1)}
$$
Here $r = |\vec{r}_1 -\vec{r}_2|$ is the separation of the
masses, and $G_\infty$ is the Newtonian gravitational constant
for $r \rightarrow \infty$.  The constants $\alpha$ and $\lambda$
characterize the strength of the new interaction (relative to
gravity), and the range of the new force.  Differentiating
$V(r)$ leads to the following expression for the force
$\vec{F}(r)$, which is what is measured in most
experiments:
$$  \everymath={\displaystyle}
\begin{array}{rcl}
    \vec{F}(r) &=& -\vec{\nabla}V(r) = -G(r) \frac{m_1m_2\hat{r}}{r^2},\\
     G(r) &=& G_\infty[1+\alpha (1+r/\lambda)e^{-r/\lambda}];
     G_o \equiv G_\infty (1+\alpha),
  \end{array}
    \eqno{(2)}
$$
We see from Eq.(2) that in the presence of a ``fifth force"
$(\alpha \not= 0)$ the usual inverse-square law breaks down. It
follows that a search for deviations from the inverse-square
law can be interpreted as a probe for new forces, and hence
of physics at the Planck scale.
The results of any test of the inverse-square law can then be
expressed in terms of an exclusion plot in the $\alpha-\lambda$
plane, as shown in Fig. 1.  (In anticipation of the ensuing
discussion, we note that tests of the inverse-square law
are also referred to as ``composition-independent" tests for
new forces.)

The stimulation for the fifth force hypothesis in 1986 came in
part from the recognition that in many specific theories the
parameter $\alpha$ in Eq.(1) is not a fundamental constant
of nature, but depends on the chemical compositions of the
test masses.  To understand how this comes about we consider
the coupling of new bosonic field to the baryon number 
$B = N+Z$, where $N$ and $Z$ denote the numbers of neutrons
and protons respectively.  The additional potential energy $V_5(r)$ arising
from the interaction of masses 1 and 2 is
$$  V_5(r) = f^2 \frac{B_1B_2}{r} e^{-r/\lambda},
    \eqno{(3)}
$$
where $f$ is a new fundamental constant.  It is straightforward
to show that the sum of Eq.(3) and the usual Newtonian potential
leads to Eq.(1) with $\alpha$ replaced by $\alpha_{12}$,
$$  \alpha_{12} = -\xi (B_1/\mu_1)(B_2/\mu_2),
   \eqno{(4)}
$$
where $\xi = f^2/G_\infty m_H^2$, $\mu_{1,2} = m_{1,2}/m_H$,
and $m_H = m(_1H^1)$.  It follows from Eqs.(1)-(4) that the
acceleration difference $\Delta \vec{a}_{12}$ of 1 and 2 towards
the Earth is given by
$$  \Delta \vec{a}_{12} = \xi (B/\mu)_\oplus
     [(B/\mu)_1 - (B/\mu)_2]\vec{\cal F} ,
    \eqno{(5)}
$$
where $\vec{\cal F}$ is the field strength of the source (in units
of acceleration), which in this case is the Earth (denoted
by $\oplus$).  For a coupling to another charge $Q$, e.g. isospin
$Q = I_z = N-Z$, one merely substitutes $B \rightarrow Q$ in
Eq.(5).
Since $\vec{\cal F}$ depends explicitly on $\lambda$ it follows
that an experimental limit on $\Delta \vec{a}_{12}$
leads to a constraint among the parameters $\xi$,
$\lambda$, and $Q$.  In practice the constraints in
the $\xi-\lambda$ plane are usually plotted for
different choices of $Q$, as in Fig. 2 for $Q=B$ and
$Q = I_z$.

Figures 1 and 2 respectively give the current (as of
March 1996) constraints on composition-independent and
composition-dependent deviations from Newtonian gravity.
In each figure the shading denotes the regions in
the $\alpha - \lambda$ or $\xi - \lambda$ plane
which are excluded by the data at the $2\sigma$
level.  We note that in each graph, the lower boundary
of the shaded region is determined by superimposing
the results of a number of different experiments.
As we discuss in Ref. [9], composition-independent
experiments achieve their maximum sensitivity for
values of $\lambda$ comparable to the dimensions of
the apparatus, and hence no single experiment can
be sensitive to all values of $\lambda$.  The situation
is somewhat different for composition-dependent
experiments but, for different reasons, it is again
necessary to rely on a collection of experiments
over different distance scales [9].

One can summarize the current experimental situation
as follows:  There is at present no compelling experimental
evidence for any deviation from the predictions of
Newtonian gravity in either composition-independent
or composition-dependent experiments.  Although there
are some anomalous results which remain to be understood,
most notably in the original E\"otv\"os experiment [10],
the preponderance of the existing experimental data is incompatible
with the presence of any new intermediate-range
or long-range forces.

We conclude this discussion by briefly 
summarizing the status of each of
the experiments or analyses in which an anomaly was
reported.

(1)  {\it E\"otv\"os, Pek\'ar, and Fekete} (1922); Ref. [10]
The EPF data were the first indication of a possible
intermediate-range composition-dependent
``fifth force".  More recent experiments with much higher
sensitivity have seen no evidence for such a force,
and hence (by implication) suggest that the EPF results
are wrong.  However, attempts to find significant
flaws in their experiment have failed, as have efforts
to explain the EPF data in terms of conventional
physics. 
There remains a slight possibility that by virtue of of its
configuration and/or
its location, the EPF experiment might have been sensitive to
a new force to which other experiments were not.
In any case, the origin and interpretation of the EPF
results remain a mystery at the present time.

(2) {\it Long} (1976); Ref. [11] This work was the
motivation for the very careful laboratory experiments of
Newman and collaborators, as well as other groups
(see, for example, Ref. [12]).
None of the more recent experiments confirm Long's
results.  Subsequent analysis by Long himself suggests
that he may have been seeing the effects of a tilt of
the floor in his laboratory as his test masses were
moved. 

(3) {\it Stacey and Tuck} (1981); Ref. [13]  This
revival of the Airy method for measuring $G_0/G_\infty$
by geophysical means initially found a result higher
than the conventional laboratory value for $G_0$. Following
the analysis of terrain bias by Bartlett and Tew
[14],
Stacey {\it et al.}
re-examined their data and concluded that the
discrepancy between their value of $G_\infty$ and $G_0$
was a consequence of having undersampled the local
gravity field at higher elevations.

(4) {\it Aronson, et al.} (1982); Ref. [15]  This
analysis of earlier Fermilab data on kaon
regeneration presented evidence for an anomalous
energy-dependence of the kaon parameters, such as
could arise from an external hypercharge field.
Since the effects reported in Ref. [15] have not been
seen in subsequent experiments, we are led to
conclude that the original data were probably biased
by some unknown (but conventional) systematic effect.
There is, however, a possibility that these
results are correct, notwithstanding the
later experiments.  This arises from the circumstance
that the data came from experiments (E-82 and
E-425) in which the kaon beam was not horizontal,
but entered the ground at a laboratory angle
$\theta_L = 8.25 \times 10^{-3}$rad (to a detector
located below ground level).  It is straightforward
to show that $\theta_L$ is related to the angle
$\theta_K$ seen by the kaons in their proper frame by
$$  \tan \theta_K  = \gamma \tan \theta_L ,
    \eqno{(6)}
$$
where $\gamma = E_K/m_K$ is the usual relativistic
factor.  For a typical kaon momentum in those experiments,
$p_K  = 70 GeV/c$, $\gamma \cong 140$ and
hence $\theta_K \cong 49^\circ$.
It follows that the incident kaons in these experiments
would have had a large component of momentum
{\it perpendicular} to the Earth, which would not
have been the case for the subsequent kaon experiments.
It can be shown that
motion of a kaon
beam perpendicular to a source of a hypercharge field
can induce an additional $\gamma$-dependence in the 
kaon parameters [16].  It is thus theoretically possible
that the ABCF results are not in conflict with the
subsequent experiments, and this could be checked in
a number of obvious ways.  Similar observations
have been made independently by Chardin.

(5) {\it Thieberger} (1987); Ref. [17]
In this experiment a hollow copper sphere floating
in a tank of water was observed to move in a direction
roughly perpendicular to the face of a cliff on
which the apparatus was situated.  Although the
reported results were compatible with the original
fifth force hypothesis, the results
of more sensitive torsion balance experiments
carried out subsequently were not.  As in the case
of the original E\"otv\"os experiment, the implication is
that Thieberger's observations can be explained in
terms of conventional physics, e.g., as a convection
effect.  

(6) {\it Hsui}, (1987); Ref. [18]. This is another determination
$G_0/G_\infty$ using the Airy method, based on
earlier data from a borehole in Michigan.
Since the original measurements
were not taken with the present objectives in mind,
it is likely that this determination of $G_0/G_\infty$
suffered from the same terrain bias that Stacey, {\it et al.}
encountered.  Moreover, a far more serious problem in
Hsui's analysis was the imprecise and very limited
knowledge of the mass distribution in the region
surrounding the borehole, which the author himself noted.

(7) {\it Boynton, et al.} (1987); [Ref. 19]
This torsion balance experiment detected a dependence
of the oscillation frequency of a composition-dipole
pendant on the orientation of the dipole relative to
a cliff.  A subsequent repetition of this experiment
by the authors using an improved pendant and apparatus
saw no effect.
Despite efforts to shield the apparatus from stray
magnetic fields, it is likely that the original
effect was due to a  small magnetic impurity in the
pendant which coupled to a residual magnetic field.

(8) {\it Eckhardt, et al.} (1988); [Ref. 20]
This was the original WTVD tower experiment in North
Carolina which saw evidence for an attractive (``sixth")
force.  
The analysis of
terrain bias by Bartlett and Tew [14]
suggested that Eckhardt, {\it et al.}, may have
undersampled the local gravity field in low-lying
regions surrounding their tower.  When the tower
results were corrected for this effect, the predicted
and observed gravitational accelerations on the tower
agreed to within errors.  A subsequent experiment by
these authors on the WABG tower in Mississippi [21] found
agreement with Newtonian gravity, as did experiments
on the Erie tower in Colorado [22]
and the BREN tower in Nevada [23].

(9) {\it Ander, et al.} (1989); [24] This
was another version of the Airy method, which used
a borehole in the Greenland icecap, and observed an
anomalous gravity gradient down the borehole.
However, this effect could not be attributed unambiguously
to a deviation from Newtonian gravity, since it
could have also arisen from unexpected
mass concentrations in the rock below the ice.


\begin{center}
{\bf REFERENCES}
\end{center}

\begin{enumerate}
\item  E. Fischbach, D. Sudarsky, A. Szafer, C. Talmadge and
S.H. Aronson, Phys. Rev. Lett. {\bf 56}, 3 (1986);
Ann. Phys. (NY) {\bf 182}, 1 (1988).
\item E.G. Adelberger, B.R. Heckel, C.W. Stubbs, and
W.F. Rogers, Annu. Rev. Nucl. and Part. Sci. {\bf 41},
269 (1991).
\item Y. Fujii, Int. J. Mod. Phys. {\bf A6}, 3505 (1991).
\item E. Fischbach and C. Talmadge, Nature {\bf 356}, 207 (1992).
\item E. Fischbach, G.T. Gillies, D.E. Krause, J.G. Schwan,
and C. Talmadge, Metrologia {\bf 29}, 215 (1992).
\item C.M. Will, ``Theory and Experiment in Gravitational
Physics", Revised Edition, (Cambridge Univ. Press, Cambridge,
1993) p. 341ff.
\item A. Franklin, ``The Rise and Fall of the Fifth Force"
(American Institute of Physics, New York 1993).
\item I. Ciufolini and J.A. Wheeler, ``Gravitation and
Inertia" (Princeton Univ. Press, Princeton, 1995) p. 91ff.
\item E. Fischbach and C. Talmadge, ``The Search for Non-Newtonian
Gravity" (American Institute of Physics, New York) in press.
\item R.v. E\"otv\"os, D. Pek\'ar, and E. Fekete, Ann. Phys.
(Leipzig) {\bf 68}, 11 (1922).
\item D.R. Long, Nature {\bf 260}, 417 (1976).
\item J.K. Hoskins, R.D. Newman, R. Spero, and J. Schultz,
Phys. Rev. {\bf D32}, 3084 (1985).
\item F.D. Stacey and G.J. Tuck, Nature {\bf 292},
230 (1981);
F.D. Stacey, G.J. Tuck, G.I. Moore, S.C. Holding,
B.D. Goodwin, and R. Zhou, Rev. Mod. Phys. {\bf 59},
157 (1987).
\item D.F. Bartlett and W.L. Tew, Phys. Rev. {\bf D40},
673 (1989), and Phys. Rev. Lett. {\bf 63}, 1531 (1989).
\item S.H. Aronson, G.J. Bock, H.Y. Cheng and
E. Fischbach, Phys. Rev. Lett. {\bf 48}, 1306 (1982).
\item D. Sudarsky, E. Fischbach, C. Talmadge,
S.H. Aronson, and H.Y. Cheng, Ann. Phys. (NY)
{\bf 207}, 103 (1991).
\item P. Thieberger, Phys. Rev. Lett. {\bf 58},
1066 (1987).
\item A.T. Hsui, Science {\bf 237}, 881 (1987).
\item P.E. Boynton, D. Crosby, P. Ekstrom, and
A. Szumilo, Phys. Rev. Lett. {\bf 59}, 1385 (1987).
\item D.H. Eckhardt, C. Jekeli, A.R. Lazarewicz,
A.J. Romaides, and R.W. Sands, Phys. Rev. Lett.
{\bf 60}, 2567 (1988).
\item A.J. Romaides, R.W. Sands, D.H. Eckhardt,
E. Fischbach, C.L. Talmadge and H.T. Kloor,
Phys. Rev. {\bf D50}, 3608 (1994).
\item C.C. Speake, {\it et al.}, Phys. Rev. Lett.
{\bf 65}, 1967 (1990).
\item J. Thomas, {\it et al.}, Phys. Rev. Lett.
{\bf 63}, 1902 (1989).
\item M. Ander, {\it et al.}, Phys. Rev. Lett.
{\bf 62}, 985 (1989).
\end{enumerate}

\begin{center}
{\bf FIGURE CAPTIONS}
\end{center}

\noindent
{\bf Figure 1.}  Constraints on $\alpha$ and $\lambda$ in Eq.(1)
implied by composition-independent experiments.  Results are
shown as of 1981 and 1996, and in each case the shaded region
is excluded at the $2\sigma$ level.
\vspace{10pt}

\noindent
{\bf Figure 2.}  Constraints on $\xi_B(a)$ and $\xi_I(b)$ as a function
of $\lambda$ from composition-dependent experiments.  
$\xi_B$ and $\xi_I$ are the coupling strengths to
$B=N+Z$ and $I_Z = N-Z$ respectively.  The shaded regions
are excluded at the $2\sigma$ level.
\vspace{20pt}

\begin{center}
{\bf R\'esum\'e}
\end{center}

\begin{quote}
En 1986 il y avait une suggestion qu'il existait
une ``cinquieme force" macroscopique dans la nature.  Cette id\'ee
a stimul\'e un nouvel int\'er$\hat{\mbox e}$t dans cette question.  On pr\'edit
detelles forces dans beaucoup de theories qui unissent la
gravit\'e avec d'autres forces connues.  On peut trouver cette
pr\'esence en cherchant des d\'eviations apparantes des
pr\'edictions de la th\'eorie de gravit\'e de Newton.  Nous r\'evisons
donc la ph\'enomenologie des recherches pour une ``cinquieme
force", et nous pr\'esentons une sommaire des r\'esultats
experimentaux courants.
\end{quote}

\clearpage
\rightline{\bf Figure 1: Constraints on $\alpha$ and $\lambda$.}
\vspace*{0.25 true in}
\centerline{\epsfxsize=6 true in\epsffile{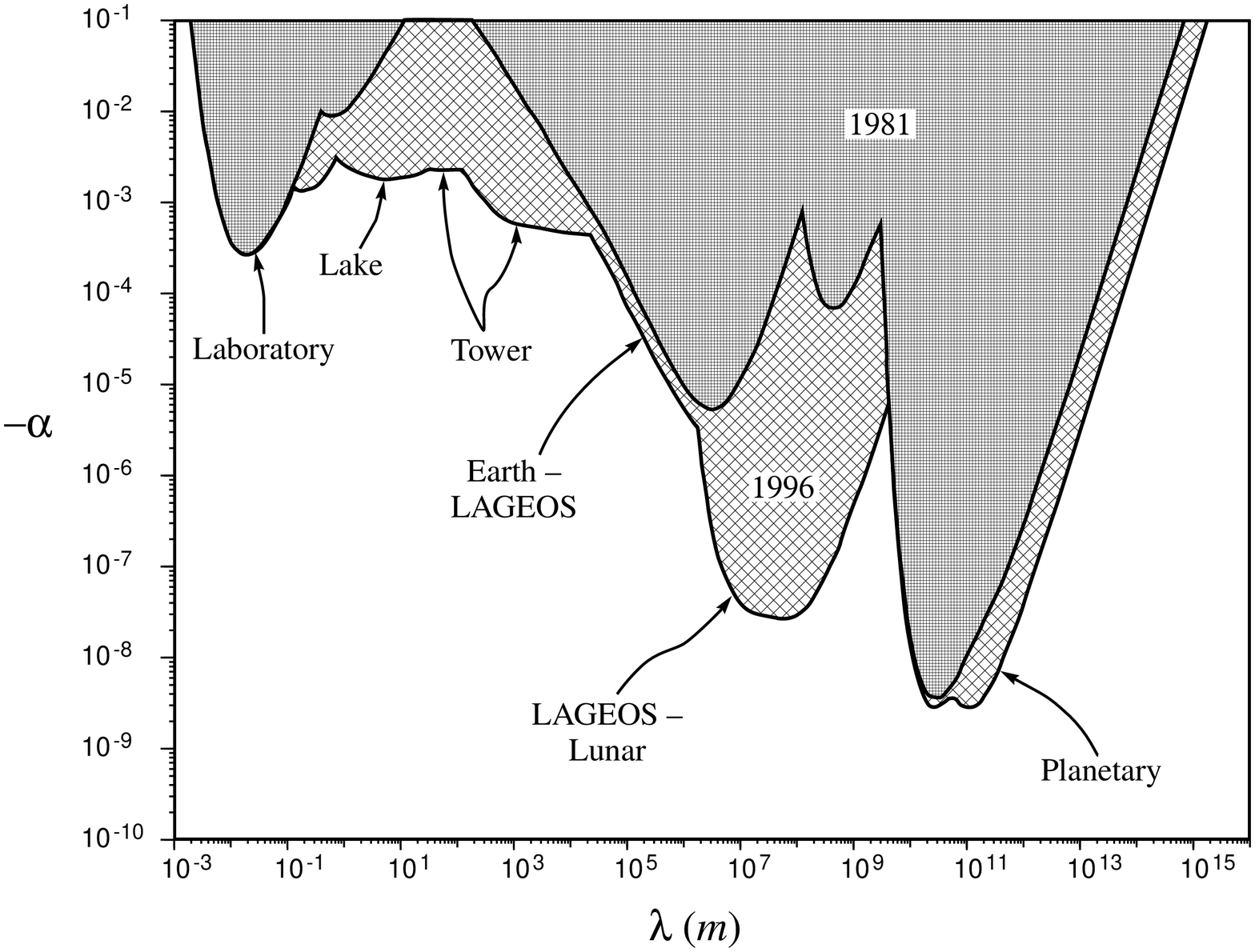}}

\clearpage
\rightline{\bf Figure 2: Constraints on $\xi_B$ and $\xi_I$.}
\vspace*{0.25 true in}
\centerline{\epsfysize=8 true in\epsffile{plot-cd.eps}}

\end{document}